\documentclass[twocolumn,prb]{revtex4}
\usepackage{graphicx}
\usepackage{dcolumn}
\usepackage{bm}
\usepackage{amsmath}

\begin{document}

\preprint{}
\title{Intrinsically  $s$-wave like property of triplet superconductors with spin-orbit coupling}
\author{T. Yokoyama, Y. Tanaka and J. Inoue }
\affiliation{Department of Applied Physics, Nagoya University, Nagoya, 464-8603, Japan%
\\
and CREST, Japan Science and Technology Corporation (JST) Nagoya, 464-8603,
Japan}
\date{\today}

\begin{abstract}
 We  have studied  a general property of unconventional superconductors with  spin-orbit coupling  by solving  the Bogoliubov-de Gennes equation  and found that  the two of four eigenfunctions for triplet pairing  coincide with those for $s$-wave pairing when a relation holds between pairing symmetry and spin-orbit coupling, indicative of an intrinsically  $s$-wave like property of triplet superconductors. 
Applying the result, we have studied the tunneling conductance in normal metal / insulator / unconventional superconductor junctions with a candidate of  pairing symmetry of CePt$_3$Si, ${\bf d}\left( {\bf k} \right) \propto {\bf \hat x}k_y {\bf  - \hat y}k_x$. The effect of Rashba spin-orbit coupling (RSOC) is taken into account. We have found that the RSOC induces a peak at the energy gap like $s$-wave superconductors in the tunneling spectrum, which stems from the intrinsically  $s$-wave like property of triplet component of CePt$_3$Si.
 As a result, the tunneling spectrum has   $s$- and $p$-wave like features. This  may serve as a tool for determing the pairing symmetry of CePt$_3$Si.
\end{abstract}

\maketitle



%

%



The recent discovery of CePt$_3$Si has attracted much attention because it is the first heavy fermion superconductor without inversion symmetry\cite{Bauer}. It is predicted that this effect  causes the Rashba type spin-orbit coupling (RSOC)\cite{Frigeri,Rashba}. Theoretically suprconductivity under broken mirror symmetry shows some novel phenomena due to the RSOC\cite{Edelstein,Gor'kov}. Howewer the effect of the other types of spin-orbit coupling (SOC), e.g., the  Dresselhaus type SOC\cite{Dresselhaus},  in superconductors without inversion symmetry is not studied well. It is important to study the general case  because the pairing symmetry and the type of SOC depend on the crystal structure. 
Frigeri et al. showed that  the pairing symmetry of  CePt$_3$Si is given by ${\bf d}\left( {\bf k} \right) \propto  {\bf \hat x}k_y {\bf  - \hat y}k_x$\cite{Frigeri} where ${\bf \hat x}$ and ${\bf  \hat y}$ are the unit vectors. This triplet pairing symmetry is, however, not conclusive\cite{Izawa}. It is desirable to calculate the physical quantities of CePt$_3$Si with candidates of its pairing symmetry and compare them with experimental results. 

All these motivate us to study the general property of unconventional superconductors (USs) with SOC. After we clarify it, we apply the result to the tunneling junctions with CePt$_3$Si and find a peak structure at the energy gap in tunneling spectrum even for the triplet pairing.  This may serve as a tool for determing the pairing symmetry of CePt$_3$Si because the tunneling spectroscopy in superconducting junctions is a powerful method to study its pairing symmetry. 

In normal metal / supercunductor (N/S) junctions Andreev reflection (AR)\cite{Andreev} is one of the most important process for low energy transport.  Taking into account the AR, Blonder, Tinkham and Klapwijk (BTK) proposed the formula for the  calculation of the tunneling conductance\cite{BTK}. This method makes it possible to clarify the energy gap profile of superconductors. It was extended to normal metal / unconventional superconductor (N/US) junctions\cite{TK,Yamashiro} in order to study the properties of unconventional superconductors. In fact, calculation or measurement of the tunneling conductance in N/US junctions are useful to study the symmetry of the pair potential of USs because  the tunneling conductance  is sensitive to the pairing symmetry due to the formation of Andreev resonant states\cite{TK,Yamashiro}. However, there is no theory considering the effect of the  SOC in the tunneling junctions. In this paper we generalize the theory in Ref.\cite{TK,Yamashiro} by incorporating  the RSOC and apply it to junctions with  a new superconductor CePt$_3$Si using the triplet pairing symmetry (${\bf d}\left( {\bf k} \right) \propto {\bf \hat x}k_y {\bf  - \hat y}k_x$). 

 We will first clarify a general property of USs with SOC by solving the Bogoliubov-de Gennes (BdG) equation. Applying the result, we calculate the tunneling conductance in normal metal / insulator / unconventional superconductor CePt$_3$Si junctions in the presence of  RSOC. We  find that the RSOC induces a peak at the energy gap in the tunneling conductance even for the triplet pairing. The result can be explained by the general property of US with SOC. Futher we estimate parameter values for the tunneling junctions with  CePt$_3$Si.  The present results may give useful information for the analysis in experiments to determine the pairing symmetry of CePt$_3$Si. Although we use a simple model for CePt$_3$Si which has a complicated band structure\cite{Samokhin}, we believe our results grasp the essence of the physics.   Below we focus on the  zero temperature regime. 


Let us start  studying  a general property of USs with SOC. Consider an effective Hamiltonian  for BdG equation with SOC. The Hamiltonian   reads
\begin{equation}
\check{H}  = \left( {\begin{array}{*{20}c}
   {\hat H\left( {\bf k} \right)} & {\hat \Delta \left( {\bf k} \right)}  \\
   { - \hat \Delta ^ *  \left( { - {\bf k}} \right)} & { - \hat H^ *  \left( { - {\bf k}} \right)}  \\
\end{array}} \right)
\end{equation}
with
$
\hat H\left( {\bf k} \right) = \xi _k  + {\bf V}\left( {\bf k} \right) \cdot {\bf \sigma }
$
, and $
\hat{ \Delta }\left( {\bf k} \right)  = i\Delta \sigma _y$ for singlet  pairing or $\hat{ \Delta } \left( {\bf k} \right) = \left( {{\bf d}\left( {\bf k} \right) \cdot {\bf \sigma }} \right)i\sigma _y 
$ for triplet pairing. Here $ \xi _k $, ${\bf k}$ and ${\bf \sigma }$ denote  electron band energy measured from the Fermi energy,  electron momentum and Pauli matrices respectively. The second term of $\hat H\left( {\bf k} \right)$, ${\bf V}\left( {\bf k} \right) \cdot {\bf \sigma }$, represents the SOC. For example, ${\bf V}\left( {\bf k} \right) \propto {\bf \hat x}k_y {\bf  - \hat y}k_x$ corresponds to the RSOC and ${\bf V}\left( {\bf k} \right) \propto {\bf \hat x}k_x {\bf  - \hat y}k_y$ does the  Dresselhaus type SOC. 
 In this work we focus on the unitary states. 
We assume $ {\bf d}\left( {\bf k} \right)\parallel {\bf V}\left( {\bf k} \right)$ where both vectors have only real number components and
$ {\bf V}\left( { - {\bf k}} \right) =  - {\bf V}\left( {\bf k} \right)$ which breaks inversion symmetry but conserves time reversal symmetry. The condition $ {\bf d}\left( {\bf k} \right)\parallel{\bf V}\left( {\bf k} \right)$ gives the highest $T_C$ of the US\cite{Frigeri}. 
The BdG equation reads
\begin{equation}
\check{H}\left( {\begin{array}{*{20}c}
   {\hat { u} _ \pm  }  \\
   {\hat{
 v} _ \pm  }  \\
\end{array}} \right) = E_ \pm  \left( {\begin{array}{*{20}c}
   {\hat{
 u} _ \pm  }  \\
   {\hat{
 v} _ \pm  }  \\
\end{array}} \right)
\end{equation}
 for electron-like quasiparticles, and
\begin{equation}
\check{H}\left( {\begin{array}{*{20}c}
   {\sigma _y \hat{
 v} _ \pm  \sigma _y }  \\
   {\sigma _y \hat{
 u} _ \pm  \sigma _y }  \\
\end{array}} \right) =  - E_ \pm  \left( {\begin{array}{*{20}c}
   {\sigma _y \hat{
 v} _ \pm  \sigma _y }  \\
   {\sigma _y \hat{
 u} _ \pm  \sigma _y }  \\
\end{array}} \right)
\end{equation}
for hole-like quasiparticles, 
with
$E_ \pm   = \sqrt {\left( {\xi _k  \pm \left| {{\bf V}\left( {\bf k} \right)} \right|} \right)^2  + \left| \Delta  \right|^2 } $, 
$
\hat{
 u} _ \pm   = u_0^ \pm  \left( {1 \pm {\bf \hat{
 V} }\left( {\bf k} \right) \cdot {\bf \sigma }} \right)
$ and
$ \hat{
 v} _ \pm   = v_0^ \pm  \frac{{\hat{
\Delta } ^\dag  }}{\left| \Delta  \right| }\left( {1 \pm {\bf \hat{
 V} }\left( {\bf k} \right) \cdot {\bf \sigma }} \right).
$
 Here
$
u_0^ \pm   = \sqrt {\frac{1}{2}\left( {1 + \frac{{\sqrt {E_ \pm  ^2  - \left| {\Delta ^2 } \right|} }}{{E_ \pm  }}} \right)}$, $ v_0^ \pm   = \sqrt {\frac{1}{2}\left( {1 - \frac{{\sqrt {E_ \pm  ^2  - \left| {\Delta ^2 } \right|} }}{{E_ \pm  }}} \right)} ,
$ 
$
\hat{\bf V}\left( {\bf k} \right) = {\bf V}\left( {\bf k} \right)/\left| {{\bf V}\left( {\bf k} \right)} \right|
$ and
 $
\left| \Delta  \right|^2  = \frac{1}{2}Tr\hat{
\Delta } \hat{
\Delta } ^\dag.  
$
 This shows that there exist the independent four eigenfunctions:  electron- and hole-like quasiparticles with the eigenvalues $E_ \pm$.

Let us discuss the property of the four eigenstates. 
For singlet pairing, we can find
\begin{equation}
 \hat{
 v} _ \pm   = v_0 \frac{{\hat{
\Delta } ^\dag  }}{\left| \Delta  \right| }\left( {1 \pm {\bf \hat{
 V} }\left( {\bf k} \right) \cdot {\bf \sigma }} \right) 
  =  - i\sigma _y v_0 \left( {1 \pm {\bf \hat{
 V} }\left( {\bf k} \right) \cdot {\bf \sigma }} \right) 
\end{equation}

with $\hat{\Delta}  = i\Delta \sigma _y$ while for triplet pairing we get 
\begin{eqnarray}
 \hat{
 v} _ \pm   &=&  - i\sigma _y v_0 \left( {{\bf \hat{
 V} }\left( {\bf k} \right) \cdot {\bf \sigma }} \right)\left( {1 \pm {\bf \hat{
 V} }\left( {\bf k} \right) \cdot {\bf \sigma }} \right) \nonumber \\ 
  &=&  \mp i\sigma _y v_0 \left( {1 \pm {\bf \hat{
 V} }\left( {\bf k} \right) \cdot {\bf \sigma }} \right)
\end{eqnarray}
with $\hat{\Delta}  =  \left( {{\bf d}\left( {\bf k} \right) \cdot {\bf \sigma }} \right)i\sigma _y$. Thus we can find that the eigenfunctions with '+' have the same form for singlet and triplet pairings. 
This indicates that the two of four eigenfunctions for triplet pairing  coincide with those for singlet pairing when the magnitude of the gap$, 
\left| \Delta  \right|$, has the same dependence on ${\bf k}$, especially with those for $s$-wave pairing when $
\left| \Delta  \right|$ is independent of ${\bf k}$. This feature gives an intrinsically  $s$-wave like property of triplet superconductors with SOC. As an example of its manifestation, we will calculate the tunneling conductance in the normal metal / unconventional superconductor (CePt$_3$Si) junctions because CePt$_3$Si is considered to satisfy  the above conditions to have an $s$-wave like property \cite{Frigeri}. 

We consider a two dimensional ballistic N/US junctions.
 The band structrure of CePt$_3$Si shows that the main contribution to the density of states stems from the $\beta$ band as shown in Ref. \cite{Samokhin}. The $\beta$ band has a three dimensional complicated structure. However, its volume is large for large $k_z$ and hence the most important part is the $\beta$ band for large $k_z$   where the dependence of the $\beta$ band on $k_z$ is weak. Therefore we focus on this part and assume the two dimensional N/US junctions as a first step. 
  The N/US interface 
located at $x=0$ (along the $y$-axis) has an infinitely
narrow insulating barrier described by the delta function $U(x)=U\delta
(x)$. We choose 
$ {\bf d}\left( {\bf k} \right) = \frac{\Delta }{{\left| {\bf k} \right|}}\left( {{\bf \hat x}k_y {\bf  - \hat y}k_x } \right)$ and 
$ {\bf V}\left( {\bf k} \right) = \lambda \left( {{\bf \hat x}k_y {\bf  - \hat y}k_x } \right) $ with  Rashba coupling constant $\lambda$. 
The eigenfunctions of the Hamiltonian are $^T \left( {u_0,\; - i\alpha _1^{ - 1} u_0,\;i\alpha _1^{ - 1} v_0,\;v_0 } \right) $, $
 ^T \left( {u_0,\;i\alpha _2^{ - 1} u_0,\;  i\alpha _2^{ - 1} v_0,\;-v_0} \right) $, $
 ^T \left( {i\alpha _1 v_0,\;v_0 ,\,\,u_0,\; - i\alpha _1 u_0} \right)$, $
 ^T \left( {i\alpha _2 v_0,\; - v_0,\,\,u_0,\;i\alpha _2 u_0} \right)$ 
with $\alpha _{1(2)}  = \frac{{k_{1(2)-} }}{{k_{1(2)} }}$, $u_0=u^+_0=u^-_0$, $v_0=v^+_0=v^-_0$,  
 $ k_1  =  - \frac{{m\lambda }}{{\hbar ^2 }} + \sqrt {\left( {\frac{{m\lambda }}{{\hbar ^2 }}} \right)^2  + k_F^2 } $,
$ k_2  =   \frac{{m\lambda }}{{\hbar ^2 }} + \sqrt {\left( {\frac{{m\lambda }}{{\hbar ^2 }}} \right)^2  + k_F^2 } $ and $k_{1(2) \pm }  = k_{1(2)} e^{ \pm i\theta _{1(2)} } $.  Here we put $E_+ =E_-$, $\theta _{1(2)}$ is an angle of the wave with wave number $k_{1(2)}$ with respect to the interface normal,  $k_F$ is Fermi wave number, and $m$ is  effective mass in US. 
Velocity operator in the $x$-direction is defined as\cite{Molenkamp}
$
v_x  = \frac{{\partial \check{H}}}{{\hbar \partial k_x }}.
$

Wave function $\psi(x)$  for $x \le 0$ (N region) is represented as 
\begin{equation}
\begin{array}{l}
 \psi \left( {x \le 0} \right) = e^{ik_y y} \left[ {e^{ik_{Fx} x} \left( {\begin{array}{*{20}c}
   {1\left( 0 \right)}  \\
   {0\left( 1 \right)}  \\
   0  \\
   0  \\
\end{array}} \right) + ae^{ik_{Fx} x} \left( {\begin{array}{*{20}c}
   0  \\
   0  \\
   1  \\
   0  \\
\end{array}} \right)} \right. \\ 
  + b\left. {e^{ik_{Fx} x} \left( {\begin{array}{*{20}c}
   0  \\
   0  \\
   0  \\
   1  \\
\end{array}} \right) + ce^{ - ik_{Fx} x} \left( {\begin{array}{*{20}c}
   1  \\
   0  \\
   0  \\
   0  \\
\end{array}} \right) + de^{ - ik_{Fx} x} \left( {\begin{array}{*{20}c}
   0  \\
   1  \\
   0  \\
   0  \\
\end{array}} \right)} \right] \\ 
 \end{array}
\end{equation}
for an injection wave in up (down) spin states with $k_{Fx}=k_F \cos \theta  x$ where $\theta$ is an angle of the wave with wave number $k_{F}$ with respect to the interface normal in the N region.
$a$ and  $b$ are AR coefficients. $c$ and $d$ are normal reflection (NR) coefficients. Because we focus on the low energy transport compared to the Fermi energy, we neglect the difference of the wave number between electron and hole. 

Similarly for $x \ge 0$ (US region) $\psi(x)$ is given by the linear conbination of the eigenfunctions. 
  Note that since the translational symmetry holds for the $y$-direction, the momenta parallel to the interface are conserved: $k_y=k_F \sin \theta  = k_1 \sin \theta _1  = k_2 \sin \theta _2 $. 

 The wave function follows the boundary conditions\cite{Molenkamp}:
\begin{eqnarray}
 \left. {\psi \left( x \right)} \right|_{x =  + 0}  = \left. {\psi \left( x \right)} \right|_{x =  - 0}  \\ 
 \left. {v_x \psi \left( x \right)} \right|_{x =  + 0}  - \left. {v_x \psi \left( x \right)} \right|_{x =  - 0}\nonumber  \\ = \frac{\hbar }{{mi}}\frac{{2mU}}{{\hbar ^2 }}\left( {\begin{array}{*{20}c}
   1 & 0 & 0 & 0  \\
   0 & 1 & 0 & 0  \\
   0 & 0 & { - 1} & 0  \\
   0 & 0 & 0 & { - 1}  \\
\end{array}} \right)\psi \left( 0 \right). 
\end{eqnarray}

Applying the BTK theory to our calculation, we obtain the dimensionless conductance  represented in the form:
\begin{equation}
\sigma _S  = \mathop \sum \limits_{ \uparrow , \downarrow } \int_{ - \frac{\pi }{2}}^{\frac{\pi }{2}} {\left[ {1 + \left| a \right|^2  + \left| b \right|^2  - \left| c \right|^2  - \left| d \right|^2 } \right]} \cos \theta d\theta.
\end{equation}

 We define the  normalized conductance as $\sigma _T =\sigma _S /\sigma _N$ where $\sigma _N$ is given by the conductance for normal states, i.e., $\sigma _S$ for $\Delta=0$ and parameters as
$\beta  = \frac{{2m\lambda }}{{\hbar ^2 k_F }}$ and 
$Z = \frac{{2mU}}{{\hbar ^2 k_F }}$. $g$ is  the effective  mass in N divided by that in US. 

\begin{figure}[htb]
\begin{center}
\scalebox{0.4}{
\includegraphics[width=16.0cm,clip]{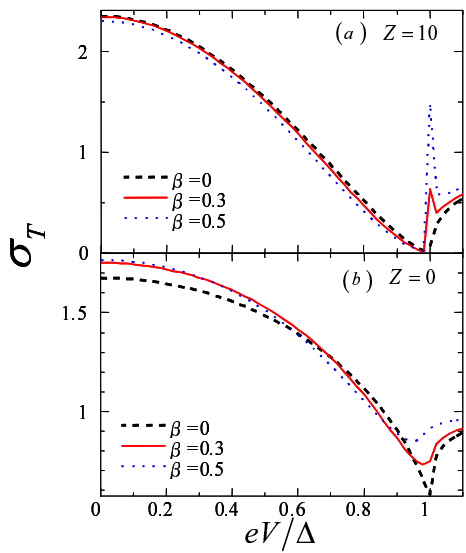}}
\end{center}
\caption{(color online) Normalized tunneling conductance with  $g=0.1$ for (a) $Z=10$  and (b) $Z=0$.} \label{f1}
\end{figure}

\begin{figure}[htb]
\begin{center}
\scalebox{0.4}{
\includegraphics[width=16.0cm,clip]{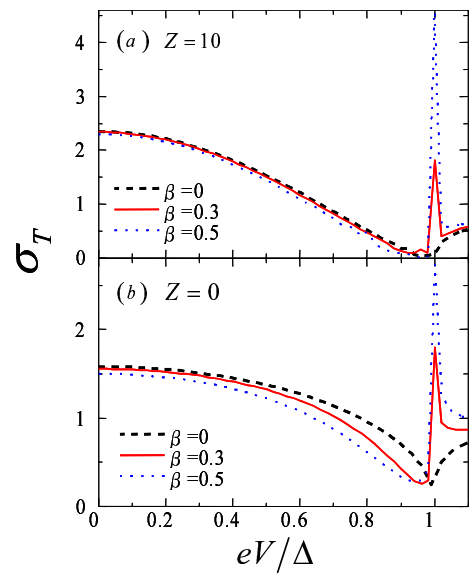}}
\end{center}
\caption{(color online) Normalized tunneling conductance with  $g=0.01$ for (a) $Z=10$  and (b) $Z=0$.} \label{f5}
\end{figure}

In the following we study the normalized tunneling conduntace $\sigma _T$ as a function of bias voltage $V$. 
For $Z=10$ and $g=0.1$, the magnitude of $\sigma _T$ at $eV=\Delta$ increases as increasing $\beta$ and finally a peak structure appears (Fig. \ref{f1} (a)).
On the other hand we can't find such a peak  for $Z=0$ and $g=0.1$ (Fig. \ref{f1} (b)). Next we will study the case of a larger effective mass in US to check its effect on the conductance.  Figure \ref{f5} shows the conductance for (a) $Z=10$ and $g=0.01$, and (b) $Z=0$ and $g=0.01$. In both cases a peak emerges at $eV=\Delta$ and becomes sharpe for large $\beta$. 

\begin{figure}[htb]
\begin{center}
\scalebox{0.4}{
\includegraphics[width=16.0cm,clip]{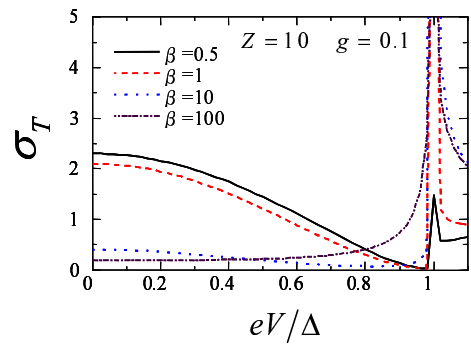}}
\end{center}
\caption{(color online) Normalized tunneling conductance for $Z=10$ and $g=0.1$.} \label{f6}
\end{figure}

Let us explain the origin of the peak at $eV=\Delta$ in Figs. \ref{f1} and \ref{f5}. A coherent peak  at $eV=\Delta$ in the tunneling conductanace also appears in N/S junctions\cite{BTK} even in the presence of RSOC\cite{Yoko}. The two of four eigenfunctions of the BdG equation with RSOC for the pair potential we consider coincide with those for $s$-wave pair potential. We can choose the electron-like and hole-like quasiparticle states with wave number $k_2$ as the two eigenfunctions. Acutually, using the expressions of $\hat{
 u} _ \pm$ and  $\hat{
 v} _ \pm$, 
we  get the eigenfunctions of the Hamiltonian for $s$-wave pairing: $^T \left( {u_0,\; - i\alpha _1^{ - 1} u_0,\;i\alpha _1^{ - 1} v_0,\;v_0 } \right) $, $
 ^T \left( {u_0,\;i\alpha _2^{ - 1} u_0,\; - i\alpha _2^{ - 1} v_0,\;v_0} \right) $, $
 ^T \left( {i\alpha _1 v_0,\;v_0 ,\,\,-u_0,\;  i\alpha _1 u_0} \right)$, $
 ^T \left( {i\alpha _2 v_0,\; - v_0,\,\,u_0,\;i\alpha _2 u_0} \right)$. This implies that US with RSOC has an intrinsically  $s$-wave like property. 
As $\beta$ increases, the contribution of the eigenstates   with wave number $k_2$ is much more dominant than that with wave number $k_1$ because $k_2$ ($k_1$) is an increasing (a decreasing) function of $\beta$. Therefore the  behavior of the conductance  becomes similar to the one for the $s$-wave pairing with increasing RSOC  in US. Note that near zero voltage the mig gap Andreev resonant states  greatly change the spectrum compared to the one for the $s$-wave junctions\cite{TK,Yamashiro}. Thus there is a qualitative difference in the conductance between two pairing states near zero voltage. For understanding the underlying physics, it is useful to check the case of very large RSOC, though it may be  unphysical.  Figure \ref{f6} displays the tunneling conductance with the same parameters in Fig. \ref{f1} (a) with very large RSOC. Apparently an $s$-wave-like tunneling spectrum emereges as increasing $\beta$, which confirms the above explanation. 

\begin{figure}[htb]
\begin{center}
\scalebox{0.4}{
\includegraphics[width=15.0cm,clip]{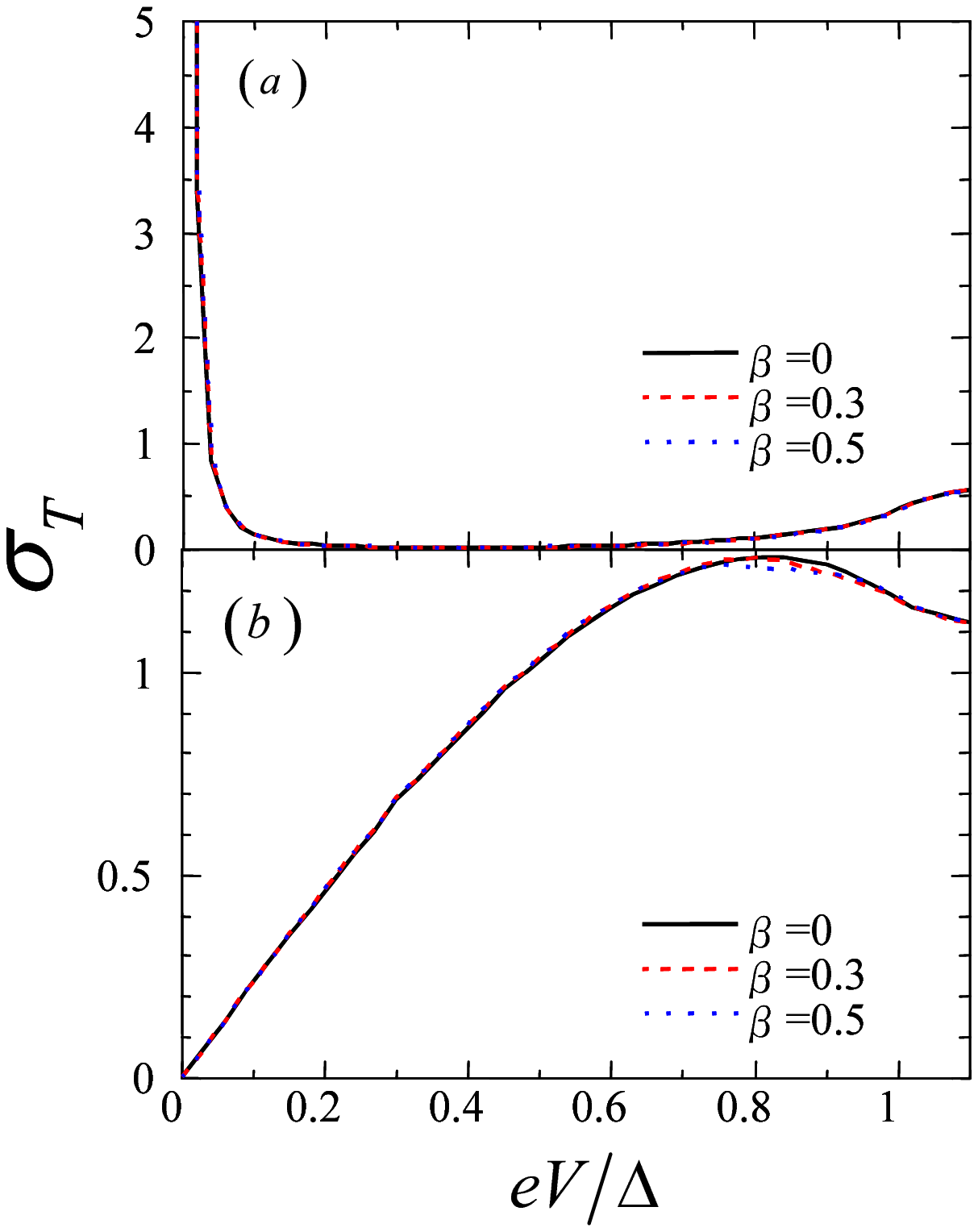}}
\end{center}
\caption{(color online) Normalized tunneling conductance for other pairing symmetries with $Z=10$ and $g=0.1$. (a) ${\bf d}\left( {\bf k} \right) = -\frac{\Delta }{{\left| {\bf k} \right|}} k_x{\bf \hat y} $ and (b) ${\bf d}\left( {\bf k} \right) = -\frac{\Delta }{{\left| {\bf k} \right|}} k_y{\bf \hat y} $.} \label{f4}
\end{figure}

Note that this feature is  unique to the pair potential considered. 
As a reference, we will consider the case of other pairing symmetries with the same parameters as in Fig. \ref{f1} (a). We choose
\begin{equation}
\hat \Delta \left( {\bf k} \right) = \left( {\begin{array}{*{20}c}
   {\Delta \frac{{k_x }}{k}} & 0  \\
   0 & {\Delta \frac{{k_x }}{k}}  \\
\end{array}} \right)
\end{equation}

in Fig. \ref{f4} (a) and
\begin{equation}
\hat \Delta \left( {\bf k} \right) = \left( {\begin{array}{*{20}c}
   {\Delta \frac{{k_y }}{k}} & 0  \\
   0 & {\Delta \frac{{k_y }}{k}}  \\
\end{array}} \right)
\end{equation}

in Fig. \ref{f4} (b).  
In both cases there is no qualitative change by the RSOC in the tunneling conduntance.

Now let us examine the parameter values suitable to CePt$_3$Si. 
In the case of CePt$_3$Si, we estimate $g \sim 0.01$ from the specific-heat measurement\cite{Bauer} and $\beta \sim 0.3$ from the band calculation\cite{Samokhin}. The corresponding results are shown by  solid curves in Fig. \ref{f5} (a) and (b), where the conductance clearly reflects the features of  $s$- and $p$-wave supercondoctors: a broad peak around zero voltage and  a peak at the energy gap.
  Recent NMR study of CePt$_3$Si showed that $1/T_1T$ of CePt$_3$Si exhibits a small peak just below $T_C$\cite{Yogi,Hayashi} like the Hebel-Slichter peak in $s$-wave superconductors\cite{Hebel}. This similarity between  CePt$_3$Si  and $s$-wave superconductors is consistent with our results. 


In summary we solved the BdG equation for US with SOC and found that  the two of four eigenfunctions for triplet pairing have the same form as those for singlet pairing under the following conditions: (i) ${\bf d}\left( {\bf k} \right)$ and ${\bf V}\left( {\bf k} \right)$ have only real components. (ii)  $ {\bf d}\left( {\bf k} \right) \parallel {\bf V}\left( {\bf k} \right)$. (iii) $ {\bf V}\left( { - {\bf k}} \right) =  - {\bf V}\left( {\bf k} \right)$. Moreover, if (iv) the magnitude of the gap, $
\left| \Delta  \right|$, is independent of ${\bf k}$, the two of four eigenfunctions for triplet pairing  coincide  with those for $s$-wave pairing. This feature gives an intrinsically  $s$-wave like property of triplet superconductors with  SOC. Note that this is essentially different from the recent work where the mixture of singlet and triplet pairing is discussed\cite{Gor'kov}. We found  that triplet superconductor itself has a singlet-like property.  As an example of its manifestation, we studied 
 the tunneling conductance in N/US  junctions with RSOC and a candidate of pairing symmetry of CePt$_3$Si. This compound is considered to satisfy the above  four conditions\cite{Frigeri}. We found that the RSOC induces a peak at the energy gap  in the tunneling conductance as found in $s$-wave superconductor junctions, which results in the $s$- and $p$-wave like features in the tunneling spectrum. This  may serve as a tool for determing the pairing symmetry of CePt$_3$Si.

%
The authors appreciate useful and fruitful discussions with S. Onari and  A. Golubov.
This work was supported by
NAREGI Nanoscience Project, the Ministry of Education, Culture,
Sports, Science and Technology, Japan, the Core Research for Evolutional
Science and Technology (CREST) of the Japan Science and Technology
Corporation (JST) and a Grant-in-Aid for the 21st Century COE "Frontiers of
Computational Science". 
%


\end{document}